# Pre-allocation Strategies of Computational Resources in Cloud Computing using Adaptive Resonance Theory-2

Dr.T. R. Gopalakrishnan Nair[1], P Jayarekha[2]

[1]Director, Research and Industry Incubation Centre(RIIC), DSI, Bangalore.

trgnair@yahoo.com

[2]Research Scholar, Dr. MGR  University  Dept. of  ISE, BMSCE, Bangalore .

Member, Multimedia   Research Group, RIIC, DSI, Bangalore.

Jayarekha2001@yahoo.co.in

## ABSTRACT

*One of the major challenges of cloud computing is the management of request-response coupling and optimal allocation strategies of computational resources for the various types of service requests.  In the normal situations the intelligence required to classify the nature and order of the request using standard methods is insufficient because the arrival of request is at a random fashion and it is meant for multiple resources with different priority order and variety. Hence, it becomes absolutely essential that we identify the trends of different request streams in every category by auto classifications and organize pre-allocation strategies in a predictive way. It calls for designs of intelligent modes of interaction between the client request and cloud computing resource manager. This paper discusses about the corresponding scheme using Adaptive Resonance Theory-2.*

## 1. INTRODUCTION

During the past few years, cloud computing has emerged as an enabling technology and it has been increasingly adopted in many areas including business, science and engineering. A cloud is an aggregation of resources/services possibly distributed and heterogeneous and operated by an autonomous administrative body which is run by a company or an organization (e.g., Amazon, Google or Microsoft). Resources in a cloud are not restricted to hardware, such as processors and storage devices, but it can also be software services or Web service in various forms of instances.

A primary driving force of the recent cloud computing paradigm is its inherent cost effectiveness[12]. The cloud computing environments are charged on the service usage, similar to many basic utilities like electricity and water. A request-response model is very appealing for both service providers and consumers and much concentration in research is required to achieve different models. As the cloud computing expands, few of the major challenges will be in the domain of fluctuating service request volume and unpredictable request arrival patterns. This unstructured demand supply scenario existing between providers and consumers can hinder the effective utilization of cloud computing environments [4]. The problem of service request scheduling in cloud computing systems is addressed in this paper. We consider a three-tier cloud structure, which consists of infrastructure vendors, service providers and consumers, and the latter two parties are of particular interest to us. Clouds are primarily driven by economics and hence the service provider aims to accommodate as many requests as possible with its objective of maximizing profit. This business interest may conflict with many of the performance parameter especially the service time. Identification of different request streams for different category and organizing the resource request pattern in a predictive way to reduce the response time, are the main objectives of this paper.

Among the multiple forms of cloud utilization, users tend to store more shared data on the cloud; the data of interest is searched with the help of service providers. One major difference existing between the cloud and the web is that the cloud data lacks the explicit link structure present in the web; this link structure plays an important role in improving the efficiency of web search.

The objects in a cluster might suggest that a particular set of objects are regularly accessed sequentially and hence it might be beneficial to prefetch them as soon as the first one is requested [13]. Further, decisions used to perform object prefetching, can in turn drive object placement decisions. For example, since objects are prefetched together, it might make sense to ensure that the objects reside on a disjoint set of hardware (nodes, routers, etc.). Furthermore, since many clients store data on the cloud, the cloud can build more robust models for clustering, prefetching, and object placement by aggregating similar access patterns for data across multiple users [13].

Hence it is required to develop a clustering and prefetching technique to effectively manage the cloud's storage. Resulting in reduction of the number of jobs being rejected and increase the profit of the service provider globally.

## 2. RELATED WORKS

### 2.1 Related work in clustering

The clustering of users based on their web access pattern is an active area of research in Web usage mining. R. Cooley et al. [2] have proposed a taxonomy of Web Mining and they present various research issues. In addition, the research in web mining is centred on the extraction and applications of clustering and prefetching. Both these issues are clearly discussed in [5]. It has been proven in this scheme that a 97.78% of prediction hierarchy is achieved.

### 2.2 Related work in prefetching

Prefetching means fetching the objects much prior to the user request arrival. There are some existing prefetching techniques, but they possess some deficiency. In this the client suffers from start-up delay for the first-time access since, prefetching action is only triggered when a client starts to access that object. However, an inefficient prefetching technique causes wastage of network resources by increasing the web traffic over the network.[7] J Yuan et al., has proposed a scheme, in which proxy servers aggressively prefetch media objects without a pattern before they are requested. They make use of servers' knowledge about access patterns to ensure the accuracy of prefetching, and have tried to minimize the prefetched data size by prefetching only the initial segments of media objects. [10] KJ Nesbit et al., has proposed a prefetching algorithm which is based on a global history buffer that holds the most recent missing addresses in FIFO order. S K Rangarajan, et al., [14] have proposed ART1 algorithm in which clustering and prediction has resulted in an accuracy of 97%. In this work we have proposed ART2 NN clustering algorithm for clustering user request arrival pattern. This cluster helps the server's agent in prefetching and making the access ready for memory, processors and most importantly storage, prior to the user request[3][7][11].

## 2.3 Related work in cloud computing

A service provider rents resources from cloud infrastructure vendors and prepares a set of services in the form of virtual machine (VM) images; the provider is then able to dynamically create instances from these VM images [4]. The underlying cloud computing infrastructure service is responsible for dispatching these instances to run on physical resources as shown in figure 1. A running instance is charged by the time it runs at a flat rate per time unit. It is in the service provider's interests to minimize the cost of using the resources offered by the cloud infrastructure vendor (i.e., resource rental costs) and maximize the revenue (specifically, net profit) generated through serving consumers' applications. From the service consumer's viewpoint, a service request for an application consisting of one or more services is sent to a provider specifying two main constraints, time and cost [12].

Although the processing (response) time of a service request can be assumed to be accurately estimated, it is most likely that its actual processing time is longer than its original estimate due primarily to due delays (e.g., queuing and/or processing) occurring on the provider's side. This time discrepancy issue is typically dealt with using service level agreements (SLAs). Scheduling strategies in this cloud computing scenario should satisfy the objectives of both parties. The specific problem addressed in this paper is the scheduling of consumers' service requests (or applications) on service instances made available by providers taking into account costs—incurred by both consumers and providers—as the most important factor.

Here we try to present, a management server and method for providing a cloud computing service at high speed and reasonable cost, are provided. The management server provides a virtual machine to a client as a computing resource. The virtual machine is multiplexed by operating multiple virtual devices on a single virtual machine. Accordingly, the demand pattern for computing resources may be predicted in advance and may be provided to a user more efficiently.

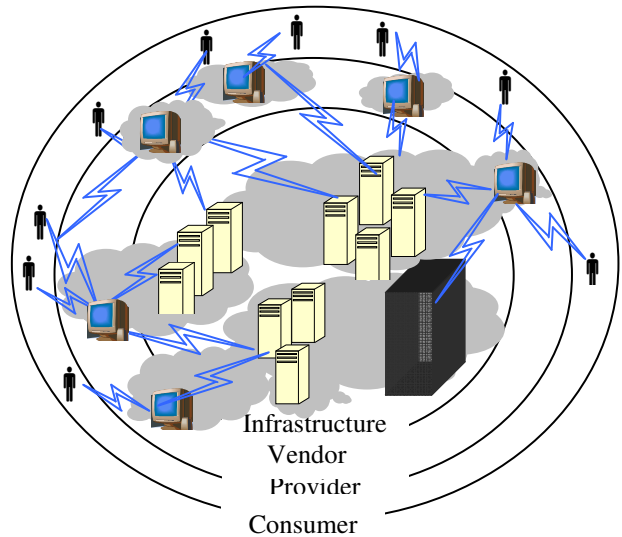

Infrastructure
Vendor
Provider
Consumer

Figure 1 Three Tier architecture of Cloud computing

## 2.4 Related work in ART2.

ART2 is an unsupervised learning and predicting algorithm derived from the resonance theory[9]. The ART network can be used for identifying patterns. This design allows the user to control the similarity between the patterns accepted by the same cluster[1]. ART2 can learn about significant new classes, yet remain stable in response to previously learned classes. Thus, it is able to meet the challenges in clustering the request arrival pattern where numerous variations are common. ART networks are configured to recognize invariant properties of a given problem domain; when presented with data pertinent to the domain, the network can categorize it on the basis of these features[6][2] This process also categorizes when distinctly different data are presented and it includes the ability to create new clusters. ART networks accommodate these requirements through interactions between different subsystems, designed to process previously encountered and unfamiliar events, respectively. We choose the ART2 neural network rather than other classifiers because it is capable of incrementally improve the numbers of clusters if needed.

ART2 networks were designed to process continuous input pattern data. A special characteristic of such networks is the plasticity that allows the system to learn new concepts and at the same time retain the stability that prevents destruction of previously learned information [6].

## 4. METHODOLOGY

### 4.1 Preprocessing the request logs

The, the log data of request arrivals are represented as <client_Id,date, requested_objects,readytime,deadline> format. We have selected a sample format of 50 clients requesting for 200 different videos.

### 4.2 Getting the Popularity Value

The popularity value is the most important parameter to get the effective prefetch operation. This popularity value may consider the long term measurement of request-frequency, which is neglected in the other algorithms. In this work the reference count value is used to get the popularity value. The reference count value is highly variable over short time scales, but this is much smoother over long time scales. This property makes the popularity value to deal with the long term measurement of request frequency.

Since the maximum and minimum value of request frequency is known during the submission of input to the ART2 system, the normalized value of the popularity can be obtained using the following transformation.

$$\delta = \frac{d - d^{min}}{d^{max} - d^{min}} \qquad (1)$$

The transformed into a range between [0 ,1].

### 4.3 Extraction of feature vectors

For clustering, we need to extract the popularity of each resource that represents the frequency of number of times that resource is requested. The pattern vector maps the access frequency of each base vector element to real values. It is of the form P ={$P_1,P_2,….P_n$} where each $P_i$ varies between 0 to 1.

| 0.8 | 0.2 | 0.2 | 0.1 | 0.3 | 0.4 | 0.5 | 0.1 | 0.3 |
|-----|-----|-----|-----|-----|-----|-----|-----|-----|

**Figure 2 Sample Pattern Vector**

Figure 2 is a sample of pattern vector generated during a session.
Each pattern vector has a real value pattern of length 200.For each session we input 50 such pattern to an ART2, since we have 50 clients.

5 PROPOSED ARCHITECTURE AND ALGORITHM

Architecture of ART2 is shown in Figure 3. It is designed for processing analog as well as binary input patterns. ART2 network module includes two main parts: attentional subsystem and orienting subsystem. Attentional subsystem preprocess analog input pattern, and then choose the best matching pattern under competitive selection rule from the input pattern prototypes[9]. Orienting subsystem carry out similarity vigilance-testing of the selective pattern prototype and trigger resonance learning and adjusting weight vectors when vigilance-testing passed, otherwise get rid of the current active node and search the other new ones. If there is no pattern prototype matching the input pattern, create a new output node to represent it. Its memory capacity can increase with the increase of learning patterns. The network allows not only off-line learning but also in an on-line learning and applying way simultaneously, that is, the learning and applying states are inseparable.

*Experimental settings*

The performance of the algorithm was thoroughly evaluated using our discrete-event cloud simulator developed in C/C++. Simulations were carried out with a diverse set of applications (i.e., various application characteristics) and settings. Owing to the large scale constitution of cloud computing environments, the number of service instances (that can be created) is assumed to be unbounded.

The total number of experiments carried out is 100 Specifically, we have generated 5 different numbers of services per application randomly selected from a uniform distribution, i.e.,U(10, 80), and 7 different simulation durations (2,000, 4,000, 8,000,12,000, 16,000, 20,000 and 30,000).

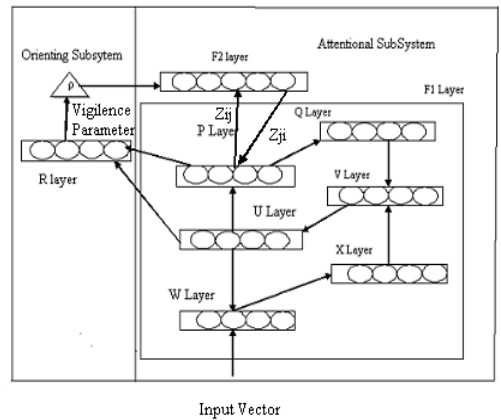

**Figure 3 The ART2 Neural networks**

The ART2 neural network algorithm used in this work is summarized below[9].

**Neural network configuration**:

The parameters required for ART2 formulation has been initially chosen :

Noise inhibition threshold:
$0 \leq \theta \leq 1$ (2)

Surveillance Parameter:
$0 \leq \rho \leq 1$ (3)
Error tolerance parameter ETP in the
F1layer:   $0 \leq ETP \leq 1$ (4)

Weight initialization of the neural
network: Top-down: $Z_{ji}(0)=0$ (5)
Bottom-up:  $Z_{ij}(0) \leq \dfrac{1}{(1-d)\sqrt{M}}$ (6)

Operation steps:

1. Initialize the sub-layer and layer outputs with zero value and set cycle counter to one.
2. Apply an input vector I to the sub-layer *W* of the F1 layer. The output of this layer is:
   $$W_i = I_i + aU_i \qquad (7)$$

3. Propagate to the *X* sub-layer:
   $$x_i = \frac{W_i}{e + \|W\|} \qquad (8)$$

4. Calculate the *V* sub-*layer* output:
   $$V_i = f(x_i) + bf(q_i) \qquad (9)$$
   In the first cycle the second term of (9) is zero once the value of $q_i$ is zero.

   The function *f(x)* is given by:

   $$f(x) = \begin{cases} \dfrac{2\theta^2}{(x^2 + \theta^2)} & 0 \leq x \leq \theta \\ x & x \geq \theta \end{cases} \qquad (10)$$

5. Compute the *U* sub-layer output:
   $$u_i = \frac{v_i}{e + \|v\|}$$
   (11)

6. Propagate the previous output to the *P* sub-layer: $p_i = u_i + dZ_{iJ}$
   (12)
   The *J* node of the F2 layer is the winner node. If F2 is inactive or if the network is in its initial configuration
   $$p_i = u_i \qquad (13)$$

7. Calculate the *Q* sub-layer output:
   $$q_i = \frac{p_i}{e + \|p\|} \qquad (14)$$

8. Repeat steps (2) to (8) until stabilizing the values in F1 layer according to *Error* (*i*) = *U*(*i*) - *U\**(*i*).
   If *Error* (*i*) ≤ *ETP*, the F1 layer is stable.

International Journal on Cloud Computing: Services and Architecture(IJCCSA),Vol.1, No.2, August 2011

9. Calculate the *R* sub-layer output:

$$r_i = \frac{u_i + cp_i}{e + \|u\| + \|cP\|}$$

(15)

10. Determine if a reset condition is indicated. If ρ (*e*) then, send a reset signal to F2, mark any active F2 node as not enable for competition, reduce to zero the cycle counter and return to the step (2). If there is no reset signal and the counter is one, the cycle counter is increased and passes to step (11). If there is no reset and the cycle counter is larger than one, then control passes to step (14), once the resonance was established.

11. Calculate the F2 layer input:

$$T_j = \sum_{i=1}^{M} P_i \, Z_{Ji} \tag{16}$$

12. Only the F2 winner node has non-zero output. Any node marked as non capable by a previous reset signal doesn't participate in the competition.

$$g(T_j) = \begin{cases} d & T_j = \max(T_k) \\ 0 & \text{otherwise} \end{cases} \tag{17}$$

13. Repeat the steps (6) to (10).

14. Update the bottom-up weights of the F2 layer winner node:

$$Z_{Ji} = \frac{u_i}{1-d} \tag{18}$$

15. Update the top-down weights of the F2 layer winner node:

$$Z_{iJ} = \frac{u_i}{1-d} \tag{19}$$

16. Remove input vector, restore inactive F2 nodes and return to the step (1) with a new input vector.

The initial values chosen for ART2 is as follows

a=10, b=10, d=0.9, c=0.1, e=0.0 θ=0.2  M = 5

Threshold value selection and methodology used

The use of several feature vectors require a special neural network, a supervised ART2 NN is used. The performance of a supervised or unsupervised ART2 NN depends on the appropriate selection of the vigilance threshold [10]. If the value of vigilance

38

threshold is near to zero, a lot of clusters will be generated, but if it is greater, then number clusters will be generated.

## 5 RESULTS AND DISCUSSION

Each request consists of the type of resources required and configuration details. A request *r* is a tuple containing at least (*n, rt, d*), where *n* specifies the number of virtual machines required; *rt* is the ready time, before which the request is not ready for execution; and *d* is the deadline for request completion. A wide range of requests to the resources can be represented by these parameters.

The users of the infrastructure run different applications with different computing requirements. Some applications need resources at particular times to meet application deadlines are called deadline-constrained applications. Whereas other applications are not strict about the time when they are given resources to execute as long as they are granted the resources required called best-effort.

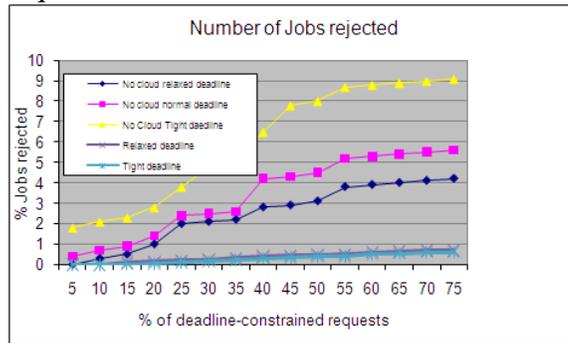

Graph 1 Comparison of job rejected

The graph 1 is a comparison for the number of jobs being rejected when no cloud and with cloud computing.ART2 model is applied for clustering and prefetching the resources. As shown in the graph the number of jobs rejected is negligible when compared with that no cloud is applied.

The service provider's interest is to minimize the cost of using the resources offered by the cloud infrastructure either provided by the vendor or in-house facility and maximize the revenue (specifically, net profit) generated through serving consumers' applications. As shown in the graph 2 cost per task completion remains constant throughout when clustering and prefetching is not applied. Along with the simulation time cost per task reduces when ART2 model is used.

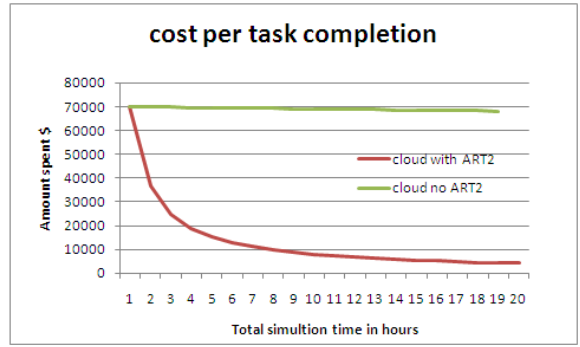

Graph 2 Comparison of cost per task completion

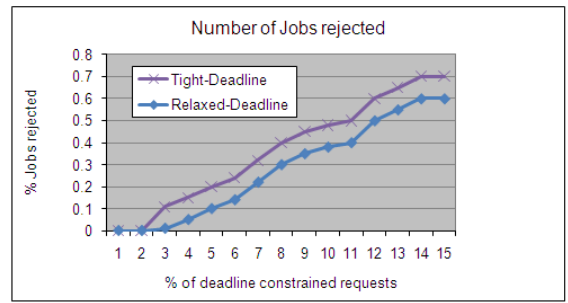

Graph 3 Comparison of job rejected

In graph 3 a comparison is presented between the Tight and Relaxed-Deadline when cloud computing provides the services by prefetching the resources. ART2 clustering technique helps in predicting the resources required hence resulting in reduction in the number of jobs being rejected even at tight deadline.

## 6 CONCLUSION

Cloud computing has the potential of becoming a revolutionary technology that changes the way service computing is performed. The principles introduced in this work are essential for the cloud computing vision to materialize to its full extent.

One of the most important benefits of Cloud computing is the ability for Cloud clients to adapt the number of resources used based on their actual use. This has great implications on cost saving as resources are not paid for when they are not used. It follows that an accurate prediction method would greatly aid a Cloud client in making its auto-scaling decisions.

Adaptive Resonance Theory-2 identifies the trends of different request streams in every category by auto classifications and organizes pre-allocation strategies in a predictive way. In the proposed design of intelligent modes of interaction between the client request and cloud computing resource manager has resulted in reduction of number of jobs being rejected and also reduction in cost per task completion.

## References

1. CARPENTER G, GROSSBERG, S , 1987, ART2: Self-Organization of Stable Category Recognition Codes for   Analog Input   Patterns,  Applied Optics, 26: p 4916:4930,

2 .COOLEY R., MOBASHER B., SRIVATSAVA J., 1997 ,Web Mining: Information and         Pattern Discovery on the World Wide Web , ICTAI'97.

3.CHRIS TSENG H, 2007,  Internet Applications with Fuzzy Logic and Neural Networks: A Survey, Journal of engineering, computing and architecture

4. Lee,Y. C.,Wang, C., Zomaya, A. Y., & Zhou B.B. (2010). *Profit-driven Service Request Scheduling in Clouds*. In Proceedings of the International Symposium on Cluster Computing and the Grid (CCGRID). Melbourne, Australia.

 5. FAUSETT, L.V. 1994,  Fundamentals of Neural Networks Architectures, Algorithms, and  applications.  New Jersey: Prentice Hall International Inc, New Jersey, p .246-287.

6. ISSAM  DAGHER , 2006 Art networks with geometrical distances  Journal of Discrete Algorithms archive    Volume 4   Issue 4   ISSN:1570-8667  Pages: 538-553.

7.JUNG J, LEE D, CHON K, 2000 Proactive  Web caching with cumulative prefetching for large multimedia Data Computer Networks – Elsevier.

8. KUMAR N, JOSHI R S  2007 Data Clustering Using Artificial Neural Networks  Proceedings of National Conference on Challenges & Opportunities in Information Technology (COIT-2007).

9. NACHEV A,  GANCHEV I 2003  Data Mining For Browsing Pattern  Weblog Data By Art2  Neural Networks International Journal Information Theories & Applications Vol.10.

10.NESBIT K J, SMITH J E, 2004 Data Cache Prefetching using a global history buffer, High Performance Computer Architecture.

11.MASSEY L  2002 Determination of clustering tendency with ART neural networks  Proceedings of recent advances in soft-computing (RASC02) .

12.Marcos Dias de Assunção, Alexandre di Costanzo, and Rajkumar Buyya. A cost-benefit analysis of using Cloud computing to extend the capacity of clusters. *Cluster Computing*, 13(3):335-347, September 2010.

13.Muniswamy-Reddy,K.K., Holland, D. A., Braun, U.,and Seltzer, M. Provenance-aware  storage systems. In proceedings of the 2006 USENIX Annual Technical Conference.

14.RANGARAJAN S K , PHOHA V V , BALAGANI K, SELMIC R R 2008 Web user caching and its application to prefetching using ART neural networks  IEEE Internet Computing, Data & Knowledge Engineering,          Vol.65 No.3, p.512-543